# Modeling of large area hot embossing

M. Worgull, K.K. Kabanemi*, J-P. Marcotte*, J.-F. Hétu*, M. Heckele

Forschungszentrum Karlsruhe, Institut für Mikrostrukturtechnik,
Postfach 36 40, 76021 Karlsruhe, Germany,
Phone: +49(7247)82-6828, fax: -4331, e-mail: Matthias.Worgull@imt.fzk.de

*Industrial Materials Institute, National Research Council of Canada

## Abstract

Today, hot embossing and injection molding belong to the established plastic molding processes in microengineering. Based on experimental findings, a variety of microstructures have been replicated so far using the processes. However, with increasing requirements regarding the embossing surface and the simultaneous decrease of the structure size down into the nanorange, increasing know-how is needed to adapt hot embossing to industrial standards. To reach this objective, a German-Canadian cooperation project has been launched to study hot embossing theoretically by a process simulation and experimentally. The present publication shall report about the first results of the simulation - the modeling and simulation of large area replication based on an eight inch microstructured mold.

## 1. Introduction

For the first time and independently of [1], embossing technology for replicating microstructures was applied by the Institute for Microstructure Technology of Forschungszentrum Karlsruhe in the early 1990s as part of the LIGA process [2]. In the course of further development, hot embossing has advanced to an independent process used apart from injection molding, injection embossing, and thermoforming. All processes mentioned have specific advantages. Consequently, they hardly compete with, but complement each other, thus covering a wide spectrum of replicated microstructures [3-6]. Process selection above all depends on the geometry of the microstructures and the surface to be patterned.

## 2. Hot Embossing

The hot embossing process is divided into four major steps:

1)  Heating of the semi-finished product to molding temperature

2)  Isothermal molding by embossing (displacement-controlled and force-controlled)

3)  Cooling of the molded part to demolding temperature, with the force being maintained

4)  Demolding of the component by opening the tool

One-sided embossing is represented schematically in Figure 1.

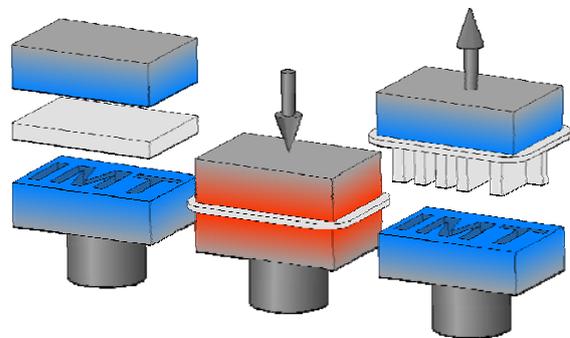

Figure 1:  The hot embossing process: Heating, molding, and demolding are the characteristic process steps. The hot embossing process is characterized by a residual layer that allows for an easy handling of the molded part.

Between the tool and substrate, a semi-finished product, i.e. a polymer foil, is positioned. Thickness of the foil exceeds the structural height of the tool.

                     

The surface area of the foil covers the structured part of the tool. The tool and substrate are heated to the polymer molding temperature under vacuum. When the constant molding temperature is reached, embossing starts. At a constant embossing rate (in the order of 1 mm/min), tool and substrate are moved towards each other until the pre-set maximum embossing force is reached. Then, relative movement between the tool and substrate is controlled by the embossing force. The force is kept constant for an additional period (packing time, holding time), the plastic material flows under constant force (packing pressure). At the same time, tool and substrate move further towards each other, while the thickness of the residual layer decreases with packing time. During this molding process, temperature remains constant. This isothermal embossing under vacuum is required to completely fill the cavities of the tool. Air inclusions or cooling during mold filling already may result in an incomplete molding of the microstructures, in particular at high aspect ratios. Upon the expiry of the packing time, cooling of the tool and substrate starts, while the embossing force is maintained. Cooling is continued until the temperature of the molded part drops below the glass transition temperature or melting point of the plastic. When the demolding temperature of the polymer is reached, the molded part is demolded from the tool by the opening movement, i.e. the relative movement between tool and substrate. Demolding only works in connection with an increased adhesion of the molded part to the substrate plate. Due to this adhesion, the demolding movement is transferred homogeneously and vertically to the molded part. Demolding is the most critical process step of hot embossing. Depending on the process parameters selected and the quality of the tool, demolding forces may vary by several factors. In extreme cases, demolding is no longer possible or the structures are destroyed during demolding. Apart from the one-sided molding described above, the process is also used for double-sided positioned embossing. The principle of the process remains the same. Instead of the substrate, however, another tool is applied. To demold the molded part from one of both tool halves, special demolding mechanisms, such as ejector pins or pressurized-air demolding, are used. For a better understanding, the schematic representation of embossing in Figure 1 is limited to the major process steps. Depending on the tool and polymer, the process and process parameters have to be adapted accordingly.

## 3. Analyzing the Hot Embossing Process

Hot embossing may be analyzed theoretically by means of a process simulation. Today, FEM simulation tools are state of the art in plastic molding. However, no simulation tool exists that satisfactorily reproduces the entire process chain of hot embossing. Complete FEM modeling of a typical LIGA mold insert using PC-based FEM systems is not yet possible due to the excessively high computational resources required to perform such an analysis. Flow behavior of polymers during embossing has already been studied for a simple microstructure [7, 8]. However, not only the individual free-standing microstructure is of interest, but also the structural field, the type of arrangement of the individual microstructures. Modeling of structural fields allows statements to be made with respect to the arrangement and mutual influence of individual structures and, thus, a tool can be designed well in advance [9-11].

The joint project covered here is aimed at analyzing the individual process steps of hot embossing, understanding related effects, and deriving improvement potentials from the findings obtained for simply structured tools. Theoretical and practical analyses focus on the demolding process, as the risk of destroying microstructures is highest during this process step. Based on simulation models, parameter studies are performed, from which conclusions are drawn with respect to optimized process parameters, above all to reduce demolding forces.

## 4. Eight inch micro structured mold

For verification of the simulation results and for the further development of design rules for large-area hot embossing, an 8-inch microstructured mold was developed and realized by micromachining (Figure 1). To replicate molds with these dimensions, a new generation of high-precision hot embossing machine was fabricated, specialized for industrial applications [12]. Advantages of the new embossing machine among others are short cycle times and an easy handling of molded parts. In combination with a sophisticated molding tool [7], basic prerequisites are provided for large-area replication. To obtain best molding conditions, shrinkage of molded parts and the demolding force measured are compared with the values predicted by simulation.

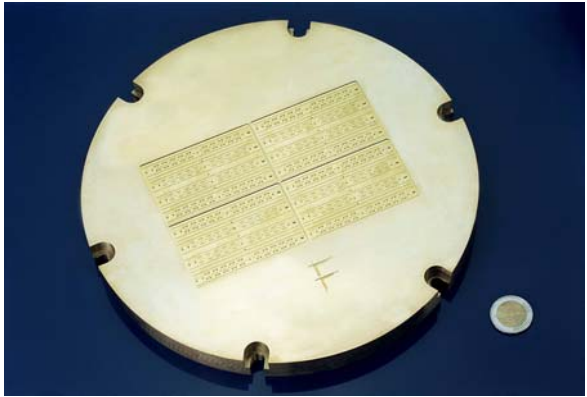

Figure 2: 8-inch microstructured brass mold. This design is used to verify the simulation results and to optimize the hot embossing process especially for large-area hot embossing.

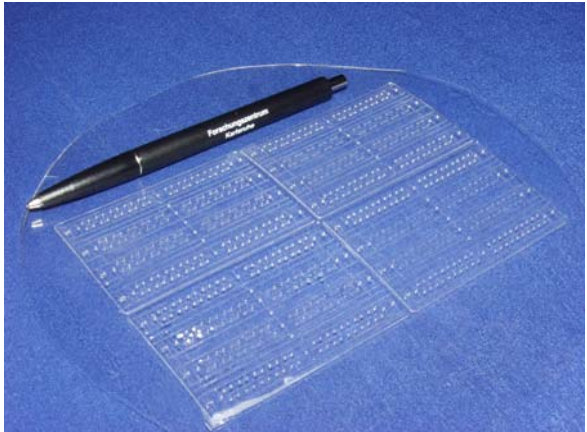

Figure 3: Molded part of the 8-inch microstructured mold.

## 5. Process Modeling

The knowledge of temperature, stresses and deformations evolutions during the molding of parts along with their dependencies on the process parameters is essential to produce low residual stress molded parts e.g. for optical applications. Although thermo-viscoelastic analysis of a single or of a few microstructures can be routinely done using commercial software, such computation is of little help to understand the influence of molding parameters on the global part shrinkage and how it affects the replication quality of the microstructures, in particular the ones located on the disc periphery. To address issues such as microstructure integrity during demolding, the analysis has to include the influence of the tooling (at the macroscopic scale) and the influence of the microstructures, which contribute to stresses, and deformations of the bulk material.

Performing such computations is a challenging task. First, they involve a very large number of small features that have to be modeled with rather fine grids. Second, the resolution of the non-linear transient thermo-viscoelasticity equations combined to contact/friction algorithms on such large grids is non trivial and has to remain accurate even though geometric feature sizes span more than 3 orders of magnitudes.

## 6. Constitutive Equations

We assume that during the cooling and demolding phases of the process that the polymer behaves as an isotropic thermorheologically simple material. Hence, stresses are related to histories of strain and temperature through appropriate relaxation functions. The latter are derived from isothermal relaxation functions by assuming time-temperature equivalence. Relaxation of the specific volume is taken into account by introducing the concept of fictive temperature, i.e., the thermodynamic equilibrium temperature is expressed as functions of the polymer thermal history, introduced by Narayanaswamy [13]. The fictive temperature is calculated by introducing a volume relaxation function in the model. The model applied here was developed by Kabanemi and Crochet and has been presented in [14] and [15]. The model will be briefly presented here but the reader is referred to the literature cited above for further details.

We denote by "$s$" and "$s_{ij}$" the spherical and deviatoric components of the stress tensor, $\sigma$, respectively, while "$e$" and "$e_{ij}$" denote the spherical and deviatoric components of the strain tensor, $\varepsilon$, respectively. For an isotropic material, we use relaxation functions $G_1$ and $G_2$ in shear and dilatation respectively, together with a modified time-scale, $\xi$. Hence, the stress-strain relation is expressed as:

$$s_{ij}(x,t) = \int_{-\infty}^{t} G_1(\xi - \xi') \frac{\partial e_{ij}(x,t')}{\partial t'} dt' \qquad (1)$$

$$s(x,t) = \int_{-\infty}^{t} G_2(\xi - \xi') \frac{\partial}{\partial t'} [e(x,t') - e_{th}(x,t')] dt' \quad (2)$$

The thermal strain, $e_{th}$, in Eq. 2 depends upon the entire temperature history of the material point and not on the temperature at time $t$ alone. The modified time scale $\xi$ at a given point, $x$(x,y,z), and at time $t$, is given by

$$\xi(x,t) = \int_{0}^{t} \Phi[T(x,\lambda)] d\lambda \qquad (3)$$

In Eq. 3, T is the temperature field and $\Phi$ the shift function often characterized by the WLF equation.

During the cooling stage, it is convenient to represent the non-equilibrium behavior of the polymer, or its structural (volume) relaxation, in terms of two variables: the actual temperature, $T$, and a fictive or structural temperature $T_f$. We assume that, at initial time $t_0$, the initial temperature $T_f(x,t_0)=T(x,t_0)$ is above the glass transition temperature $T_g$, and that the specific volume $v(t_0)$ of the material is given. Following Narayanaswamy [13], the specific volume, $v(t)$, may be given by an integral equation of the type:

$$e_{th}(\mathbf{x},t) = \int_{T(\mathbf{x},t_0)}^{T_f(\mathbf{x},t)} \alpha_l(T')dT' + \int_{T_f(\mathbf{x},t)}^{T(\mathbf{x},t)} \alpha_g(T')dT' \quad (10)$$

where $\alpha_l$ and $\alpha_g$ are the linear thermal expansion coefficients in the liquid and glassy states, respectively. Finally, one needs a constitutive equation for the evolution of the fictive temperature $T_f$ for which we use the following model

$$T_f(\mathbf{x},t) = T(\mathbf{x},t) - \int_0^t M_v(\xi-\xi')\frac{\partial T(\mathbf{x},t')}{\partial t'}dt' \quad (11)$$

where $M_v$ is a volume relaxation function. In the liquid region, the relaxation is fast, and one obtains $T_f(x,t)=T(x,t)$. On the other hand, when the material is quenched below the glass transition temperature from temperature $T(t=0)=T_0$ above $T_g$, the relaxation does not exist and one obtains $T_f=T_0$. In this work, the same relaxation function is used for both $M_v$ and $\varphi$.

## 7. Contact and Friction Conditions

In the present formulation we assume that the mold insert and the substrate behave as rigid bodies. We consider the case of a unilateral contact, which involves no penetration between the two bodies and is modeled with the Signorini conditions. We start the analysis by considering that a full contact zone, $\Gamma$, is developed at the entire interface between the polymer sheet and the mold insert. The first condition to satisfy is the non-penetration condition. If $\mathbf{n}$ denotes the interior normal to the mold insert then the non-penetration boundary conditions on $\Gamma$ can be written as

$$u_n \leq 0 \quad (12)$$
$$f_n \leq 0 \quad (13)$$
$$f_n \cdot u_n = 0 \quad (14)$$

Here $u_n$ is the normal displacement at the interface and $f_n$ is the normal contact force. The first inequality, Eq. 12, represents the kinematics condition of no penetration of the contact surface. The second inequality, Eq. 13, is the static condition of compressive or zero normal tractions. The third equation, Eq. 14, states that there is zero work done by the normal contact stress, i.e., the normal contact stresses exist only at the nodes where the polymer sheet is in contact with the rigid mold insert.

We model the frictional contact between the viscoelastic body and the mold insert with a Coulomb's law of dry friction, written as

$$|f_t| < \mu_s|f_n| \quad (15)$$

Here $\mu_s$ is a static friction coefficient associated with the stick friction constraint and $f_t$ represents the tangential force on the contact boundary $\Gamma$. This static version of the Coulomb's law states that the tangential shear cannot exceed the maximal frictional resistance. When the strict inequality, $|f_t| < \mu_s|f_n|$, holds the surface of the polymer adheres to the mold insert and is in the so-called stick state, and when the inequality, $|f_t| \geq \mu_s|f_n|$, holds there is relative sliding, the so-called slip state. The Coulomb's friction law is modified to include a dynamic friction coefficient, $\mu_d$, associated with the slip condition when the static constraint is violated. In that case the inequality, $\mu_d \leq \mu_s$, holds. It follows from the above analysis that the interface mold/polymer is divided in three zones, which are not known a priori and are part of the problem: stick, slip and no contact or gap.

We use the penalty method to enforce contact constraints and a regularization technique to obtain estimates for the normal contact force, $f_n$, and the tangential frictional traction, $f_t$, as follows

$$f_n = -\lambda_n u_n \quad (16)$$
$$f_t = -\lambda_t u_t \quad (17)$$

where $\lambda_n > 0$ and $\lambda_t > 0$ are normal and tangential penalty parameters, respectively. The result is a solution to the contact problem that allows small violations of the contact constraints in order to estimate the direction and magnitude of the actual tractions.

## 8. Governing Equations

Based on these assumptions, the mechanical problem of frictional contact of the viscoelastic deformable polymer can be formulated as follows:

*Find a temperature field T, a displacement field **u**, and a stress field $\sigma$ such that:*

$$\rho c_p \frac{\partial T}{\partial t} = \nabla \cdot (k\nabla T) \quad on\ \Omega$$

with:

$$T(\mathbf{x},0) = T_{initial}$$

$$k\frac{\partial T}{\partial n} = h(T - T_{mold}) \quad on\ \Gamma$$

and

$$\nabla \cdot \boldsymbol{\sigma} + \boldsymbol{f} = 0 \quad in \ \Omega \ ,$$

$$u_n \leq 0 \ , \ f_n \leq 0 \ , \ f_n \cdot u_n = 0 \ , \text{on} \ \Gamma,$$

with $|f_t| < \mu_s |f_n| \quad \Rightarrow u_t = 0,$

$\quad |f_t| \geq \mu_s |f_n| \quad \Rightarrow \exists \lambda_t > 0,$

such that $f_t = -\lambda_t u_t .$        (18)

where $\Gamma$ represents the part-mold interface.

## 9. Numerical Methods

In order to solve the equations expressed in Eq. (18), a standard Galerkin finite element formulation is used. Non-linearities associated with material constitutive equations and contact/friction boundary conditions are linearized using a Picard method. See reference [15] for more details on the numerical implementation.

The numerical methods just described have been implemented into dFEMwork, IMI's finite element software toolkit for large scale parallel computing. This toolkit provides distributed data structures such as matrices and vectors, parallel preconditioned Krylov iterative solvers, dynamic data redistribution. Domain decomposition was performed using ParMETIS parallel graph partitioner [16]. Algebraic systems were solved using ILU(0) conjugate gradient iterative method [17].

## 10. Application to the 8-inch mold

The described approach has been applied to predict temperature, stress and deformation fields during the cooling and demolding of the polymer part. The mesh generation was performed using ANSYS AI Environment. Because of the very large number of features and because of the nature of the equations that have to be solved, mesh generation proved to be challenging. Because of the symmetry of the part, only one quarter was meshed and simulated. Figure 4 shows the surface mesh. Linear 4-node tetrahedral elements are used. The finite element mesh, consisted of 1 256 122 elements, was exported and used as input into the parallel finite element solver described earlier.

The polymer used is a PMMA (BASF Lucryl G77Q11), whose master curve at 110 °C is given in Figure 5. We have chosen a relaxation spectrum characterized by four relaxation times. The part is initially at 183°C and is cooled with convection on both top and bottom.

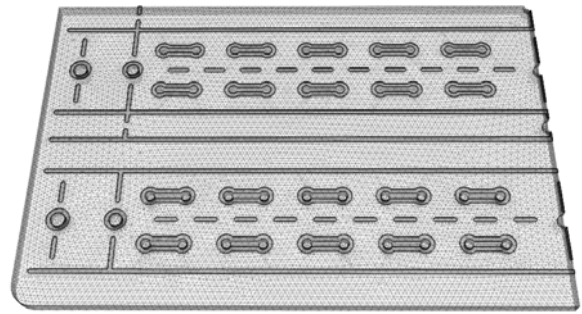

Figure 4: 3D Meshed microstructured molded part. Note that, only one quarter of only one of the 4 sections of the 8-inch microstructured part have been meshed. This figure represents the surface mesh. The mesh has 1 256 122 elements.

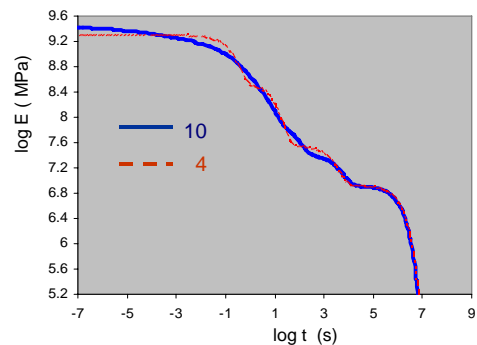

Figure 5: Relaxation modulus as a function of time for a PMMA at 110 °C using 4 and 10 relaxation times.

The temperature distribution, at t=3 s, during cooling in the mold is shown in Figure 6. Large temperature gradients in gap-wise direction are predicted.

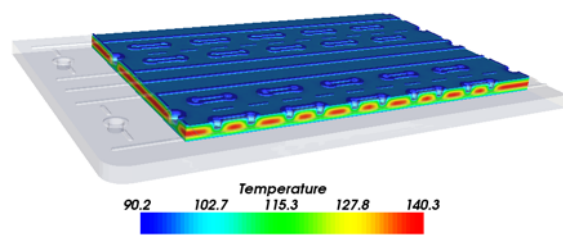

Figure 6 : Three-dimensional temperature distribution (t=3 s) in the molded part.

The stress field and final shape of microstructures at the end of cooling inside the mold are shown in Figures 7 and 8, respectively. These results correspond to the viscoelastic calculation without friction. As can be seen from Figure 7, large in-plane tensile stresses are predicted on the different microstructures compared to the small stresses in regions without microstructures. These results are due to both the rate of cooling that induces large temperature gradients and the complexity of the microstructure geometry (mold insert), which prevents the part to deform freely inside the mold. We also observe that residual stresses are mainly localized around corners in all microstructures. These results will have an impact on both the shape of microstructures and the quality of the replication. This observation is highlighted in Figure 8, where the deformed microstructures are shown (deformation have been amplified with a scale factor of 25). In this figure we clearly see the effect of the geometry (ribs and hollows) on the deformation inside the mold.

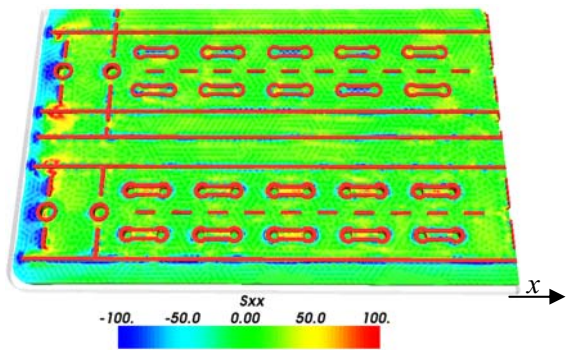

Figure 7: In-plane stress distribution ($s_{xx}$) in the microstructures inside the mold, at the end of cooling process.

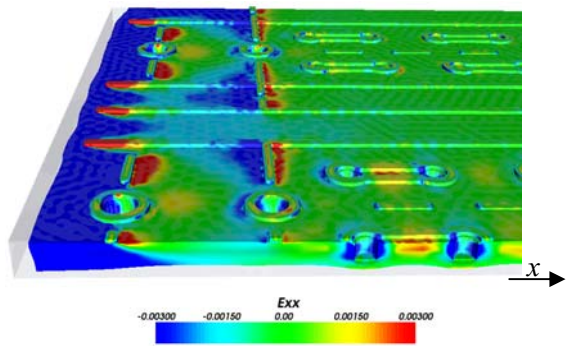

Figure 8: Zoomed section, with a cut through feature, of the strain distribution ($\varepsilon_{xx}$) after cooling inside the mold with deformed molded part (with scale factor 25).

## Acknowledgement

This work has been funded by the NRC Helmholtz Science and Technology Fund and the German Ministry of Education and Research, BMBF (project 01SF0201/7).